\documentclass[12pt]{article}

\usepackage{graphicx}
\usepackage{color}
\usepackage[colorlinks=true, urlcolor=blue, linkcolor=blue, 
citecolor=blue]{hyperref}
\usepackage{cite}

\newcommand{\beq}{\begin{equation}}
\newcommand{\eeq}{\end{equation}}
\newcommand{\beqa}{\begin{eqnarray}}
\newcommand{\eeqa}{\end{eqnarray}}
\newcommand{\beqar}{\begin{eqnarray*}}
\newcommand{\eeqar}{\end{eqnarray*}}

\begin{tiny}

\end{tiny} 

\begin{document}
\thispagestyle{empty}
$\,$

\vspace{30pt}

\begin{center}

\textbf{\Large TeV gravity searches}

\vspace{50pt}
Jos\'e I.~Illana, Manuel~Masip 
\vspace{16pt}

\textit{CAFPE and Departamento de F{\'\i}sica Te\'orica y del Cosmos}\\
\textit{Universidad de Granada, E-18071 Granada, Spain}\\
\vspace{16pt}

\texttt{jillana,masip@ugr.es}

\end{center}

\vspace{30pt}

\date{\today}

\begin{abstract}
In scenarios with extra dimensions the gravitational interaction may 
become strong at TeV energies. This could modify the $\nu$$N$
cross section and imply distinct signals at 
neutrino telescopes. In particular, cosmogenic neutrinos of 
$E\approx 10^9$ GeV could experience frequent interactions with 
matter where they
lose a very small fraction of their energy. We define a consistent
model of  strong gravity at the TeV scale with just one extra dimension and 
a first Kaluza-Klein excitation of the graviton of mass around 1 GeV.
We describe the collisions at transplanckian energies 
(multigraviton exchange, graviton emission and black hole formation)
as well as the possible signature of these processes at km$^3$ telescopes 
and their impact in cosmogenic neutrino searches.
\end{abstract}

\vspace{2cm}

\noindent
Prepared for the book {\it Particle Physics with Neutrino Telescopes}, 
C. P\'erez de los Heros, editor (World Scientific)

\newpage
\section{Introduction}\label{ra_sec1}
The hierarchy problem, namely, how to make consistent 
a quantum field theory that includes very different scales, has defined
the model building in particle physics during the past four decades.
In units of the Planck mass, $M_P=1.2\times 10^{19}$ GeV, 
the electroweak (EW)
scale ($m_{h}^{2}\approx 10^{-34} M_P^2$) and the vacuum energy
density ($\Lambda\approx 10^{-120} M_P^4$) have extremely small values.
These two scales are free parameters in our theory, but they
include ${\cal O}(1)$ quantum corrections that
require a large fine tuning in order to reproduce the values that we see.
The usual strategy to explain {\it naturally} the fine tuning 
in $m_{h}^{2}$ had been to complete the standard model 
with new symmetries (TeV particles that cancel the quantum corrections
to $m_h$) and/or new dynamics (the Higgs as a composite of a new interaction 
that becomes strong at the TeV). In both cases the result
may be an EW scale with only logarithmic sensitivity to the ultraviolet (UV) physics.

In this context, it is difficult to overstate the impact on the community
of the 1998 paper on extra dimensions and TeV gravity by Arkani-Hamed,
Dimopoulos and Dvali (ADD) \cite{ArkaniHamed:1998rs}. All the 
basic ingredients in their
analysis were already
known: there were previous proposals of compact dimensions 
with radius $R\approx (1$ TeV)$^{-1}$ to
break supersymmetry \cite{Antoniadis:1990ew} or 
with $R\approx (10^{12}$ GeV)$^{-1}$ in M-theory to
lower $M_{\rm string}$ to $M_{\rm GUT}\approx 10^{16}$ GeV
\cite{Horava:1996ma}, and it had even been shown 
how different fields of the same theory can 
live in a different number of 
dimensions (D-branes) \cite{Polchinski:1998rr}. However, ADD
realized that the fundamental scale of gravity was not necessarily 
$M_P$, and that it could be as low as 1 TeV. In this case, obviously, the 
hierarchy problem introduced by the EW scale would disappear. The model by ADD
has 3 basic parameters that are related by $M_P$: the fundamental scale 
of gravity ($M_D$), the number
of compact dimensions ($n$) and their radius ($R$). One year later 
Randall and Sundrum (RS) \cite{Randall:1999ee,Gherghetta:2010cj} 
generalized the framework with 
a new parameter, a higher dimensional curvature ($k$) that was
zero in the ADD model.

Since then, the models with extra dimensions have revealed a
very rich phenomenology. Not only they are able to accommodate 
hierarchies, they also seem to provide an 
alternative ({\it holographic}) description of strongly coupled 4-dimensional 
theories \cite{Maldacena:1997re,ArkaniHamed:2000ds}, 
which opens unlimited possibilities for model building. 
If the LHC confirms the absence of new physics 
below 1 TeV we will learn that nature does not deal with the hierarchy
problem the way we assumed, but this will not diminish the relevance
of extra dimensions. On one hand, they remind us that
a more fundamental scale may appear anywhere, so they represent a stimulus
for the exploration of higher energies. On the other hand, they may be
used to explain the apparent fine tuning in the free parameters of a model 
({\it e.g.}, the small Yukawa couplings 
required in the standard model, specially if neutrinos have a 
Dirac nature \cite{Dienes:1998sb}) or to define consistent models 
of TeV physics able to explain 
{\it any} experimental anomaly ({\it e.g.}, a large forward-backward asymmetry
in $t \bar t$ production at the Tevatron \cite{Barcelo:2011vk}).

\begin{figure}[t]
\centerline{\includegraphics[scale=1]{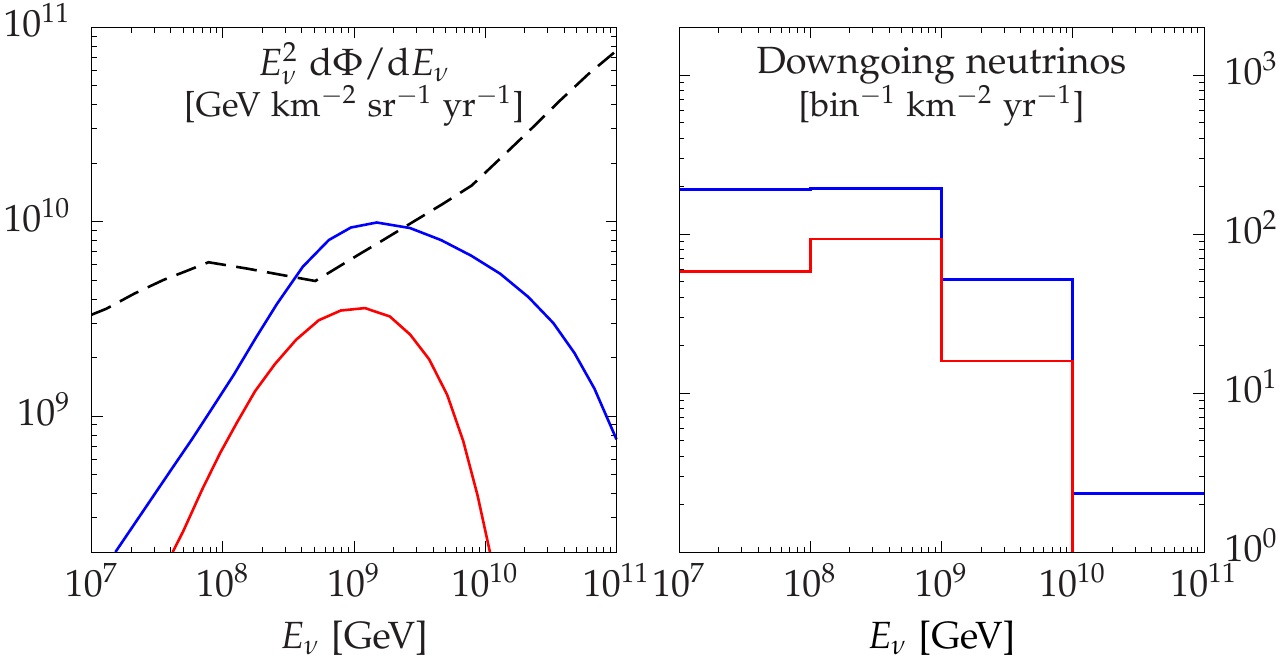}}
\caption{Cosmogenic neutrino fluxes (minimal \cite{Fodor:2003ph} and higher \cite{Ahlers:2010fw}) with 90\% CL upper limits from IceCube (dashed \cite{Aartsen:2017mau}) (left) and number of downgoing neutrinos in each energy bin (right).
}
\label{fig1}
\end{figure}

Strong TeV gravity is particularly relevant 
for the physics of ultrahigh energy neutrinos. 
What makes neutrinos special is that they only have weak
interactions, implying that the relative effect of the new physics could be
more important than for quarks and charged leptons. In particular, cosmogenic neutrinos
appear when $10^{10}$--$10^{11}$ GeV cosmic rays propagate and interact 
inelastically with the 2.7~K cosmic microwave background 
radiation \cite{Semikoz:2003wv}. 
This cosmogenic $\nu$ flux is certainly
there (see Fig.~\ref{fig1}) \cite{Ahlers:2012rz}, and it implies 
a few tens of neutrinos of energy around
$10^9$ GeV reaching the Earth per km$^2$, unit of solid angle and year
\cite{Fodor:2003ph,Ahlers:2010fw}.
In the collision of such neutrinos with a nucleon the center of mass (c.o.m.) energy  
is $\sqrt{s}=\sqrt{2m_NE}\approx 45$ TeV, well above the scale 
explored at colliders. 
Since the standard model interaction length in ice for a $10^9$ GeV neutrino is 
around 1000 km, cosmogenic $\nu$s could (should!) be detected in the near future.
Moreover, their absence in experiments like ANITA \cite{Deaconu:2017eyy}, 
LUNASKA \cite{James:2009rc} or IceCube-Gen2 \cite{Aartsen:2014njl}
could mean that new physics may be hiding them. We will show that
this could be the case if, at energies above a threshold around $10^7$ GeV,
neutrinos experience transplanckian interactions with matter.

The plan of this chapter is as follows. First we will discuss in some detail 
the basic ideas
of TeV gravity through extra dimensions and will define a consistent set up.
Then we will obtain the neutrino--nucleon cross section in that
framework and 
discuss its validity at high energies. In particular, we will 
argue that at $s\gg M_D^2$ the result is independent of the UV details
of the theory, {\it i.e.}, of how gravity is embedded in a consistent 
quantum theory.
Finally we will discuss the possible signal of these scenarios at
large-scale neutrino telescopes.

\section{Extra dimensions: circles, orbifolds and curvature}
We have mentioned that the usual strategy to solve the hierarchy problem
had been to search for a mechanism that keeps $m_h$ much smaller
than $M_P$. Extra dimensions provide the opposite approach: they
explain why the Planck mass is so much larger than the EW scale. Let us see how
it works.
The Planck mass is defined by Newton's law for the gravitational force
between two masses, $M$ and $m$, separated by a distance $r$ ($c=1=\hbar$):
\beq
F(r)=-G_N \, {Mm\over r^2}
\label{Newton}
\eeq
with
$G_N={1/M_P^2} \equiv 1/(8\pi \bar M_P^2)$.
This dependence with the distance reflects Gauss' law: the flux of field lines
through a Gaussian surface of radius $r$ around $M$ is constant, 
so its density {\it dilutes} like $1/r^2$. If gravity 
were propagating in just two spacial dimensions instead of three, then Newton's law
would go like $1/r$, whereas 1-dimensional gravity would imply a constant 
($r$ independent) force. Now suppose that gravity propagates in 2 dimensions,
but that the second one is compact and has a length $L$, as given in Fig.~2.
\begin{figure}
\centerline{\includegraphics[angle=0,width=7cm]{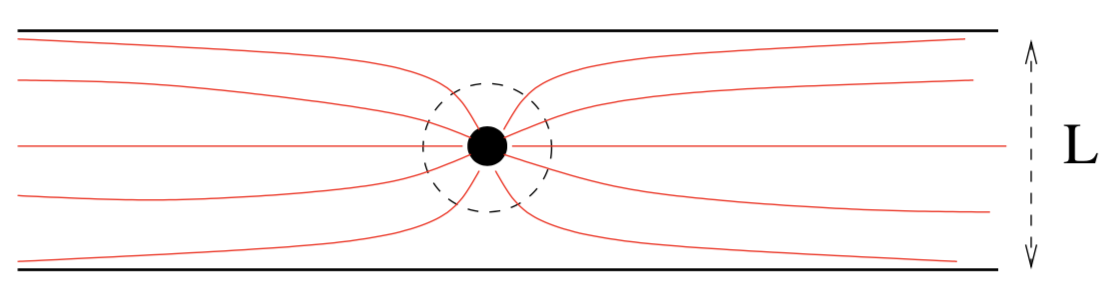}}
\caption{Field lines in a 2-dimensional space where 1 dimension is compact and of length $L$.}
\label{fig2}
\end{figure}
At distances much shorter than $L$ the field lines are insensitive to the fact that
this dimension is compact, and gravity will be purely 2-dim. At larger
distances, however, the field lines do not dilute any longer 
and gravity becomes 1-dim, {\it i.e.}, there is a constant density of field
lines. If we consider the usual 4-dimensional space plus one extra dimension compactified on
a circle $S^1$ of radius $R$, at $r\gg R$ we will verify the usual 
Newton's law, while
at $r\ll R$ we will have
\beq
F(r)=-G_5 \, {Mm\over r^3}\,.
\label{5D}
\eeq
It is important to notice that the 5-dimensional gravitational 
constant has now dimensions of
$M^{-3}$:
\beq
G_5\equiv {1\over 8 \pi\,\bar M_5^3}\,.
\eeq
$\bar M_5$ is the fundamental scale in this (4+1)-dimensional theory, whereas 
$\bar M_P$  
is an effective scale that appears in
long-distance interactions. Matching (\ref{Newton}) and (\ref{5D}) 
 at $r\approx L=2\pi R$ we find how these two scales relate:
\beq 
{G_5\over L}= G_N\hspace{0.5cm}{\rm or}\hspace{0.5cm}\bar M_P^2=\bar M_5^3\,L\,.
\eeq
For $n$ extra dimensions ($D=4+n$) defining a torus $T^n=(S^1)^n$ with (common) radii $R$, 
an analogous argument gives 
\beq 
{G_D\over V}= G_N\hspace{0.5cm}{\rm or}\hspace{0.5cm}\bar M_P^2=
\bar M_D^{2+n}\,(2\pi R)^n\,,
\eeq
where $G_D=1/(8\pi \bar M_D^{2+n})$.\footnote{Warning: 
the scale most frequently used in
the literature is $M_D=\bar M_D(2\pi)^{n\over 2+n}$ \cite{Giudice:1998ck}, 
which does not coincide
with the original $M_*$ in \cite{ArkaniHamed:1998rs} 
either.} The fact is that the 
gravitational interaction grows at short distances $r<R$ faster than in $D=4$, and it
becomes strong
at a scale $\bar M_D$ that (varying $R$) may take values between $m_h$ and 
$M_P$.

\subsection{Kaluza-Klein modes}
Another important concept in higher-dimensional theories is that of Kaluza-Klein
(KK) mode. It may be instructive to see how KK excitations 
appear in the simplest set up, a 
5-dimensional complex scalar field with the extra dimension 
compactified on a circle. Let us label the coordinates
$x^M=(x^\mu,x^5)$ with $x^5=y$ and use the metric $\eta_{MN}={\rm Diag}(-1,1,1,1,1)$.
The action for the free field $\Phi(x^M)$ is
\beq
S_5=-\int {\rm d}^5 x \; \partial_M\Phi^\dagger \,\partial^M\Phi \,,
\label{S5S}
\eeq
where ${\rm d}^5 x = {\rm d}^4x\,{\rm d}y$ and $[\Phi]=E^{3/2}$. Since in $S^1$
 we
identify $y$ with $y+2\pi R$, we can expand the $y$ dependence 
\beq
\Phi(x^\mu,y)={1\over \sqrt{2\pi R}} \sum_{n=-\infty}^{+\infty}\phi^{(n)}\! (x^\mu)\,
e^{i {ny\over R}}\,.
\label{phi5}
\eeq
Using this expansion in (\ref{S5S}) and integrating over $y$ we get 
$S_5=S_4^{(0)}+S_4^{(n)}$, with 
\beqa
S_4^{(0)} &=&-\int {\rm d}^4 x \, \partial_\mu\phi^{(0)\dagger} \partial^\mu\phi^{(0)}\,, 
\nonumber \\
S_4^{(n)} &=&-\int {\rm d}^4 x  
\sum_{n\not= 0} \left( \partial_\mu\phi^{(n)\dagger} \partial^\mu\phi^{(n)} +
 \left({n\over R}\right)^2\,  \phi^{(n)\dagger}\phi^{(n)} \right) \,.
\eeqa
We have traded the field dependence on $y$ by a tower of 4-dim
KK modes of
mass a multiple of $m_c\equiv 1/R$. This mass is nothing but the 
quantized momentum
$p_y$ of  $\Phi$ along the compact dimension: notice that
$\partial_5\Phi^\dagger \partial^5\Phi$ becomes the mass term in the 
4-dimensional action and that each KK level
includes two modes, reflecting that $p_y$ may be positive or negative.
The KK masses for compactification on a $n$-dimensional torus are
$m^2_{n_1,n_2,...}=(n_1^2+n_2^2+...)/R^2$, which correspond to a momentum
$\pm n_i/R$ along each extra dimension.

The KK expansion for the graviton is a bit more involved. The metric is
a symmetric tensor, and in 5 dimensions its  
fluctuations ($g_{MN}=\eta_{MN}+h_{MN}$) 
have $15$ independent components:
\beq
h_{MN}=h_{\mu\nu}\oplus h_{\mu 5}\oplus h_{55}\,.
\eeq
The 5-dimensional Einstein-Hilbert action admits
local transformations that may be used to eliminate ten of them, leaving five physical
polarizations: two in $h_{\mu\nu}$, two in $h_{\mu 5}$ and $h_{55}$.
The zero modes of these 5-dimensional fields will
define the 4-dimensional graviton plus a vector and a real scalar field. 
As for the massive ($n\not= 0$)
modes, $h_{\mu 5}^{(n)}$ and $h_{5 5}^{(n)}$ are {\it eaten} by $h_{\mu \nu}^{(n)}$ to
define a KK tower of spin-two massive fields, each one with 5 physical polarizations. 
In more than five dimensions there are additional KK towers of scalar and 
vector fields \cite{Giudice:1998ck}. 

\subsection{Orbifolds}
\begin{figure}
\centerline{\includegraphics[width=0.9\linewidth]{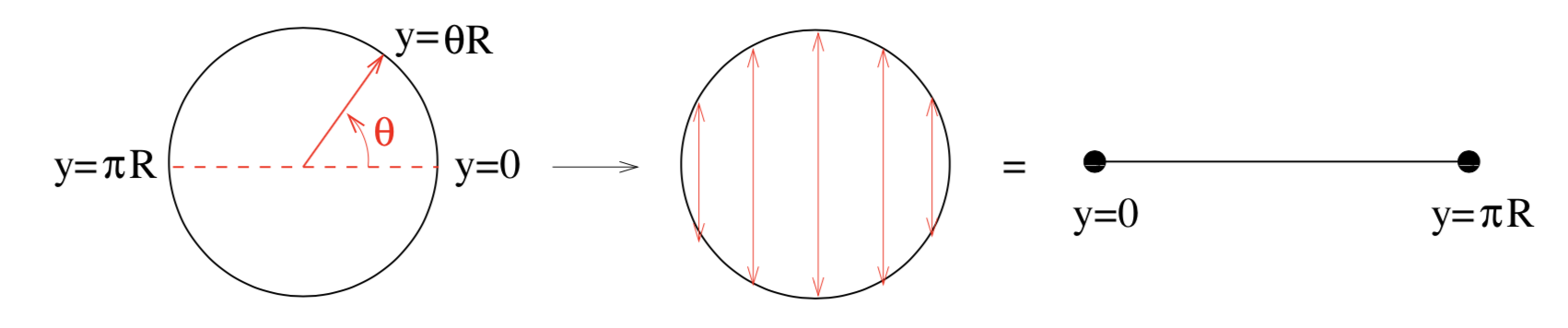}}
\caption{Orbifold $S^1/Z_2$.}
\label{fig3}
\end{figure}

Compactification of an extra dimension on $S^1$ faces a main problem: when we 
reduce fermions to four dimensions they are always vector-like (Dirac) fields; the theory does
not admit chiral fermions. The basic reason is that $\gamma_5$ is now part of
the Dirac algebra,
\beq
\{ \gamma^M,\gamma^N\} =2\eta^{MN}\,,
\eeq
and the boosts along the extra dimension will change the chirality of the fermion. The 
solution to this is to change the compactification space: instead of a differentiable manifold we will
use an orbifold whose singularities will
break Lorentz invariance along $y$. The orbifold $S^1/Z_2$
is obtained from the circle $-\pi R \le y \le \pi R$ by identifying $y\to -y$, as
shown in Fig.~\ref{fig3}.

What kind of 5-dimensional fields can live on this orbifold? Consider a scalar
field $\Phi(x,y)$; an obvious guess would be that
only those 5-dimensional fields in $S^1$ with $\Phi(x,-y)=\Phi(x,y)$ survive in
$S^1/Z_2$. There is, however, a second
possibility: fields with $\Phi(x,-y)=-\Phi(x,y)$. The reason is that a field by 
itself is not
physical, only the action is. If the 5-dimensional action on $S^1$ contains only even 
powers of $\Phi$, then the action will be invariant under $y\to -y$ even if
 $\Phi(x,-y)=-\Phi(x,y)$. In other words,
we may include the global $Z_2$ symmetry $\Phi\to -\Phi$ in the $Z_2$ modding
and obtain a consistent theory on the orbifold. This is more clear if one defines
directly the theory on $S^1/Z_2$ instead of deforming the parent $S^1$ theory. 
Take a 5-dimensional real scalar field:
\beq
S_5=-\int {\rm d}^5 x \; \left(
{1\over 2} \,\partial_M\Phi \,\partial^M \Phi +V(\Phi) \right)\,.
\label{S5}
\eeq
Its motion is obtained imposing $\delta S=0$, with
\beqa
\delta S &=& - \int {\rm d}^5 x \; \left(
{\partial {\cal L}\over \partial \Phi} \,\delta \Phi + 
{\partial {\cal L}\over \partial (\partial_M\Phi)} \,\delta (\partial_M \Phi) \right)\nonumber \\
&=& - \int {\rm d}^5 x \; \left(
{\partial V \over \partial \Phi}\, \delta \Phi + 
\partial^M \!\Phi\, \partial_M (\delta \Phi) \right)\nonumber \\
&=&  \int {\rm d}^4 x \int _0^{\pi R} \!\! {\rm d} y \, \left[
\left( \partial_M \partial^M \Phi - {\partial V \over \partial \Phi}\right) \delta \Phi 
-\partial_M \! \left( \partial^M\! \Phi\; \delta \Phi \right) \right]\,.
\eeqa
The first term equal zero gives the 5-dimensional equations of motion, and 
the second term implies the boundary conditions (BCs) $\Phi\to 0$ at $x^\mu\!\to \infty$ and 
$(\delta \Phi\;\partial_5\Phi)_{y=0,\pi R}=0$. At each 4-dimensional {\it brane} this can be satisfied 
in two different ways:
\beqa
\partial_5 \Phi&=&0\hspace{1cm}{\rm Neumann,}\nonumber \\
\Phi&=&0\hspace{1cm}{\rm Dirichlet}\,.
\eeqa
The KK expansion of $\Phi$ will be affected by these boundary conditions.
If $\Phi$ satisfies the Neumann boundary condition at $y=0$ and $y=\pi R$ we have
\beq
\Phi_+(x,y) = {1\over \sqrt{\pi R}}\,\phi^{(0)}_+ + 
{ \sqrt{2\over\pi R}}\,\sum_{n=1}^\infty \phi^{(n)}_+ \cos{ny\over R}\,,
\eeq
whereas a field with Dirichlet boundary conditions must be expanded
\beq
\Phi_-(x,y) =
{\sqrt{2\over \pi R}}\,\sum_{n=1}^\infty \phi^{(n)}_- \sin{ny\over R}\,.
\eeq
The KK modes above have been normalized so that upon integration of
the extra dimension they have 4-dimensional canonical kinetic terms. Notice that in both cases
the boundary conditions imposed by the orbifold eliminate (project out) 
half the KK tower living in the circle and, most important, 
for a Dirichlet boundary condition there is {\it not} a zero mode. 
In a general theory, the compactification on $S^1/Z_2$
implies a $Z_2$ parity that can be used to break
gauge symmetries or to define chiral fermions. More precisely, it is easy to see that
the $A_\mu$ and $A_5$ components in a vector field or the $\Psi_L$ and 
$\Psi_R$ spinors in a 5-dimensional fermion have opposite parities; as a consequence 
the KK tower of one of those
fields ({\it e.g.}, $A_5$ and $\Psi_R$) will not include a massless mode.

\subsection{Curvature}
The 5-dimensional space defined in the previous section has a non-trivial topology but
a flat metric. RS found a very interesting deformation of this space: they 
introduced
constant energy densities in the bulk ($\Lambda_5=-6k^2 \bar M_5^3$) 
and the two 4-dimensional branes ($\Lambda_0=-\Lambda_{\pi R}=\Lambda_5/k$),
\beqa
S\supset \int {\rm d}^4 x \,{\rm d} y \left[ \sqrt{-g} \left({1\over 2} \,\bar M_5^3 {\cal R}+
\Lambda_5 \right) + 
 \sqrt{-g_0}\; \delta(y) \,\Lambda_0 \right. \nonumber \\
+ \,\sqrt{-g_{\pi R}}\; \delta(y-\pi R)\, \Lambda_{\pi R}\bigg]\,,
\label{EHaction}
\eeqa
so that the space becomes
a 5-dimensional slice of anti-de Sitter (AdS$_5$). The tuning of these three energy densities is 
equivalent to the requirement of a vanishing cosmological constant in a 4-dimensional theory. 
The Einstein equations for this action are solved by the metric\footnote{
Contrary to the original RS model, we place the infrared (IR) brane at $y=0$ and
the UV brane at $y=\pi R$. This set up, proposed in \cite{Giudice:2004mg}, 
can be easily obtained from
the usual one in \cite{Randall:1999ee} by redefining $y\to \pi R - y$.}
\beq
{\rm d} s^2= e^{2ky} \eta_{\mu\nu} \, {\rm d} x^\mu {\rm d} x^\nu + {\rm d} y^2\,.
\eeq
Integrating $y$ in Eq.~(\ref{EHaction}) one finds 
a 4-dimensional Einstein-Hilbert action with \cite{Giudice:2004mg}
\beq
\bar M_P^2 = \bar M_5^3 \int_0^{\pi R} {\rm d} y \; e^{2ky} =  {\bar M_5^3\over 2k}
\left( e^{2k\pi R} -1 \right)\,.
\eeq
This relation between
the fundamental scale 
$\bar M_5$ and $\bar M_P$ generalizes the one in ADD. If $k\to 0$, 
taking $e^{2k\pi R}\approx 1 + 2k\pi R$
we have the ADD relation $\bar M_P^2 = \bar M_5^3 L$, 
but if $kR>1$ then $e^{2k\pi R}$ may be much larger than $1$ 
(implying $\bar M_5 \ll \bar M_P$) even if all the scales 
($k$, $R^{-1}$ and $\bar M_5$) are similar.
This scenario allows for exponentially different scales to coexist at different points
of the fifth dimension, as the {\it natural} scale $\bar M_5$ at $y=0$ is 
{\it blue shifted} by the metric to $\bar M_P$ at the UV ($y=\pi R$) brane.

The quantum fluctuations of the metric in Eq.~(\ref{EHaction}) will 
include the massless graviton $h_{\mu\nu}^{(0)}$,
the radion $h_{55}^{(0)}$ and a KK tower of massive gravitons $h_{\mu\nu}^{(n)}$,
whereas 
the zero mode of $h_{\mu 5}$ is projected out by the orbifold boundary conditions and the 
massive modes $(h_{\mu 5}^{(n)}$ and $h_{55}^{(n)})$
are eaten by the KK gravitons. The 
equations of motion ($\delta S=0$) for $h_{\mu\nu}(x,y)$ on this 
{\it warped} orbifold are: 
\beq
\partial_\rho\partial^\rho \, h_{\mu\nu}+e^{-2ky} \, \partial_5 \! \left( e^{4ky} \partial _5 h_{\mu\nu} \right) = 0\,, 
\eeq
with $\partial_5 h_{\mu\nu}=0$ at $y=0,\pi R$. The KK expansion is then
\beq
h_{\mu\nu}(x,y)={1\over \sqrt{\pi R}} \sum_{n=0}^\infty h_{\mu\nu}^{(n)}(x)\, 
f^{(n)}(y)\,,
\label{KKexp}
\eeq
where $f^{(n)}$ is an eigenfunction of $p_y^2$ 
({\it i.e.}, $\partial_y^2 f^{(n)}= m_n^2 f^{(n)}$) with normalization
\beq
\int_0^{\pi R} {\rm d}y\;e^{2ky} f^{(n)} f^{(m)} =\pi R \,\delta_{m n}\,.
\eeq
This means that $f^{(n)}$ satisfies 
\beqa
\left[ e^{-2ky} \, {{\rm d}\over {\rm d} y} \left( e^{4ky} {{\rm d}\over {\rm d} y}  \right)
+m_n^2 \right]  f^{(n)}(y) &=& 0\,, \nonumber \\
\partial_y f^{(n)}(y)\bigg|_{y=0,\pi R} &=& 0\,.
\eeqa
The solutions can be given in terms of Bessel functions:
\beq 
f^{(n)}(y)={z_n^2\over N_n}\left[ J_2(z_n)+c_n Y_2(z_n)\right]\,,
\eeq
where $z_n\equiv m_n e^{-ky}/k$ and the constants $N_n$ and $c_n$ are fixed
by the orthonormality conditions and the boundary condition at $y=0$ \cite{Giudice:2004mg}. 
The mass eigenvalues
$m_n$ are then obtained from the boundary condition at the UV brane. For $k<1/R$ it results
the usual KK spectrum on the circle, $m_n\approx n/R$, but with half the modes (the
orbifold projects out the other half). For 
$k>1/R$, however, the graviton masses ($n>0$) become
\beq
m_n\approx \left( n+{1\over 4} \right) \pi k\,. 
\eeq 
So it is the curvature $k$ and not $1/R$ what defines the mass of the first
excitation and the gap between KK modes. Finally, the interaction of these
gravitons with a field at the IR ($y=0$) brane is deduced from
\beq
S\supset -{1\over \bar M_5^{3/2}}
 \int {\rm d}^5 x \; e^{4ky}\; h_{\mu\nu}(x,y)\; T^{\mu\nu}(x)\,\delta(y)\,,
\eeq
which upon integration on $y$ implies the 4-dimensional Lagrangian 
\beq
{\cal L} = - T^{\mu\nu} \left( {1\over \bar M_P}
 \, h_{\mu\nu}^{(0)}+ \sum_{n=1}^{\infty} {1\over \Lambda_n}
 \, h_{\mu\nu}^{(n)} \right)\,.
\label{KKint}
\eeq
When the curvature is negligible one obtains
$\Lambda_n^{-1}= \sqrt{2}/\bar M_P$ (the flat result on the orbifold), whereas
for $k<1/R$ the
curvature pushes the KK modes towards the IR brane and increases their
couplings to $\Lambda_n^{-1}= e^{k\pi R}/\bar M_P$ \cite{Giudice:2004mg}.
It is interesting to notice that, if we define 
\beq
m_c \equiv \left\{ 
\begin{array}{l} 
R^{-1}\hspace{0.5cm} {\rm if}\;\; k<R^{-1}\\
\pi k \hspace{0.75cm} {\rm if}\;\; k>R^{-1},
\end{array} \right. 
\eeq
then 
we can write an expression for the scale $\Lambda_n$ that works 
in both limits:
\beq
\Lambda_n^2 = {\bar M_5^3 \,\pi \over 2 \,m_c}= {M_5^3  \over 4 \,m_c}\,,
\eeq
where we have used $M_5^3=2\pi \bar M_5^3$.
This common expression for the scale $\Lambda_n$ setting the strength of the massive
graviton interactions 
lets us understand easily 
the gravitational potential both  in the flat and the warped cases
\cite{Illana:2014bda}.
Consider first an ADD model ($k=0$) with one extra dimension
compactified on the orbifold and a fundamental scale $\bar M_5=1$ TeV. 
To reproduce 
$\bar M_P=2.4\times 10^{18}$ GeV we need $L=\bar M_P^2/\bar M_5^3=7.4$ AU,
or $m_c=5.4\times 10^{-19}$ eV. At distances $r>7.4$ AU the
potential created by each KK graviton is
suppressed by a Yukawa factor of $e^{-2 m r}$ and we can neglect its
contribution to 
the usual (4-dimensional) Newton potential. At $r<L$, however, all the KK gravitons
of mass $m_n<1/r$ will be {\it active}. Since their multiplicity is 
$r^{-1}/m_c $ and they couple with the same
strength, their effect in the potential is to change
\beq
{1\over \bar M_P^2} \to {4 \,m_c \over M_5^3} \times {r^{-1}\over R^{-1}} 
={4\over M_5^3 \;r}\,.
\eeq
Now consider a 5-dimensional set up with  also $\bar M_5=1$ TeV
but a non-zero value of $k>R^{-1}$,
for example, $k=0.3$ GeV, which implies $m_c=1$ GeV. 
The 4-dimensional Newton potential dictated by the massless graviton 
will now extend down to distances 
$r$ of order $1/m_c=0.2\;{\rm fm}$, whereas at smaller distances the
effect of the KK excitations gives
\beq
{1\over \bar M_P^2} \to {4 \,m_c \over M_5^3} \times {r^{-1}\over \pi k} =
{4\over M_5^3 \;r}\,.
\eeq
At these shorter distances the flat and the warped results coincide.
The density of KK modes in the warped case is lower when
$k>R^{-1}$, but their coupling to matter is stronger and both effects compensate: 
a single graviton of mass
$m_n$ in the warped model creates the same potential as 
all the
KK modes with mass between $m_{n-1}$ and $m_n$ in the flat case. 
The curvature in a RS model provides then an extra parameter that just rises the 
mass of the 
first KK mode but  implies, at distances $r<m_c$, the same gravitational interaction as the
{\it simpler} ADD scenario.

\section{Transplanckian collisions}
A $2\to 2$ collision between light particles with the mediator in the
$t$-channel is characterized by two kinematical variables: the c.o.m energy
$\sqrt{s}$ and the momentum transfer $q=\sqrt{-t}$, with $s$ and $t$ the
usual Mandelstam variables.\footnote{In terms of the metric defined in the 
previous section, $s=-(p_1+p_2)^2$ and  $t=-(p_1-p_3)^2$.}
In the weakly coupled regime $q$ defines the typical 
impact parameter
in the collision, $b\approx 1/q$, and its value will determine the scattering angle in
the c.o.m frame: a forward collision corresponds to 
a long-distance interaction 
with a small value of $q$, whereas larger values of $q$ up to $\sqrt{s}$ imply short
distance processes and larger angles. Take a
neutrino of energy $E$ scattering off a particle of mass $m$ initially at rest. 
The neutrino will lose in the collision a fraction $y$ of its energy (the {\it inelasticity}), 
whereas the target particle will gain an energy $yE$. 
It is easy to see that $q$ also determines
the value of $y$: $y=q^2/s$. Therefore, one may refer to a low-$q$ process
as a forward or a soft collision.

In the 5-dimensional model of TeV gravity outlined in the previous section, we will be 
interested in transplanckian collisions of $\sqrt{s}> M_5$. An obvious objection would
be whether such collisions would require a UV complete theory of gravity. This, however,
is not the case: all the processes of interest will be dominated by long-distance
interactions that are insensitive to the UV physics.
Let us discuss this in some detail.

\subsection{Eikonal amplitude}
Graviton-mediated interactions are better understood in 
impact parameter space \cite{Amati:1993tb,
Emparan:2001ce,Giudice:2001ce,Emparan:2001kf,Illana:2004qc,Illana:2005pu}.
At distances $b$ 
larger than the inverse mass of the first KK graviton the scattering amplitude 
should be frozen, as the massless graviton couples with a strength
suppressed by $M_P$ and the KK modes do not reach beyond $b\approx m_c^{-1}$. 
As the distance decreases more KK gravitons become active,
and the process is described by an eikonalized amplitude
${\cal A}_{\rm eik}(s,t)$ that includes the infinite set of ladder and cross-ladder
diagrams. Basically, ${\cal A}_{\rm eik}(s,t)$ is the exponentiation of the Born
amplitude in impact parameter space: 
\beq
{\cal A}_{\rm eik}(s,t)={2 s\over i}
\int {\rm d}^2b\; e^{i\mathbf{q}\cdot\mathbf{b}}\;
\left(e^{i\chi (s,b)}-1\right)\;,
\label{eq4}
\eeq
where $\chi (s,b)$ is the eikonal phase,
$\mathbf{b}$ spans the 2-dimensional impact parameter space
and $t\approx -q_\perp^2$.
The eikonal process will be reliable as far as
the integral is dominated by $b>r_H$ (see below), and 
it reduces to the Born amplitude for a small eikonal phase. 
In the transplanckian
regime it is also 
independent of the spin of the colliding particles.
The phase $\chi (s,b)$
can be deduced from the Fourier transform to
impact parameter space of ${\cal A}_{\rm Born}(s,t)$:
\beq
\chi(s,b) = \frac{1}{2s}\int\frac{d^2q_\perp}{(2\pi)^2}
{\rm e}^{-i{\bf q}\cdot{\bf b}}{\cal A}_{\rm Born}(s,-q_\perp^2)\ .
\label{eikphase}
\eeq
Our Born amplitude comes from the $t$-channel exchange
of the KK graviton tower:
\beqa
{\cal A}_{\rm Born}(s,t)&=&-{4 m_c s^2 \over M_5^{3}}
\sum_{n=1}^\infty {1\over t-(n m_c )^2} \nonumber \\
&=& {2 \pi s^2 \over M_5^{3} q} 
\left( \coth { \pi q\over m_c} 
- {m_c\over \pi q} \right)
\,,
\label{eq5}
\eeqa
where $q=\sqrt{-t}$ and we have not included 
the contribution of the massless
graviton (with a much smaller coupling than the massive
modes if $m_c>R^{-1}$). At $q>m_c$ we have
$\tilde {\cal A}_{\rm Born}\approx {2 \pi^2 s^2 / (M_5^{3} q)}$ whereas
at $q\to 0$ the amplitude becomes
${\cal A}_{\rm Born}(s,0)= {2 \pi s^2 /( 3 M_5^{3} m_c)}$. In our calculation
of  ${\cal A}_{\rm eik}$ 
we will use the first expression for all the values $q$ and  will then 
correct the result for low $q$.
The eikonal phase is in that case
\beq
\chi(s,b) = {s\over 2 M_5^3} \int_0^\infty {\rm d} q \;
 J_0(qb)
\equiv \frac{b_c}{b} \,,
\eeq
with 
$b_c={s/(2 M_5^3)}$.
The divergence in the eikonal phase at short distances ($b=0$) does
not affect the amplitude in Eq.~(\ref{eq4}): the contributions 
from the region 
$b\ll b_c$ are quickly oscillating and tend 
to cancel. This is also the basic reason why ${\cal A}_{\rm eik}$ is 
insensitive to the UV completion of gravity.

The eikonal amplitude can then be written 
\beq
\tilde {\cal A}_{\rm eik}(s,t)=4\pi s\, b_c^2\; F_1(b_c q)\;,
\label{eik2}
\eeq
where the tilde indicates that the expression is not valid
at $q\to 0$ and  
\beq
F_1(u)=-i
\int_0^\infty {\rm d}v\;v\; J_0(uv)
\left( e^{iv^{-n}} -1 \right)\;.
\label{eq7}
\eeq
It is easy to see that  the integral defining ${\cal A}_{\rm eik}(s,t)$ 
in Eq.~(\ref{eq4}) 
is dominated by a saddle point at $b_s=(q^2 b_c)^{-1}$ if $q>b_c^{-1}$
and by $b\approx q^{-1}$ for  $q<b_c^{-1}$.
The modulus of the complex function above is
$|F_1(u)|\approx 1/\sqrt{1.57u^3+u^2}$, and 
$\tilde {\cal A}_{\rm eik}$  can be
 corrected at low $q$ by reintroducing the factor that
we took from ${\cal A}_{\rm Born}$ in Eq.~(\ref{eq5}):
\beq
{\cal A}_{\rm eik}(s,t)=4\pi s\, b_c^2\; F_1 (b_c q)
\left(\coth {\pi q\over m_c}- {m_c\over \pi q} \right) 
\;.
\label{eikonal}
\eeq
\begin{figure}
\begin{center}
\includegraphics[scale=1.]{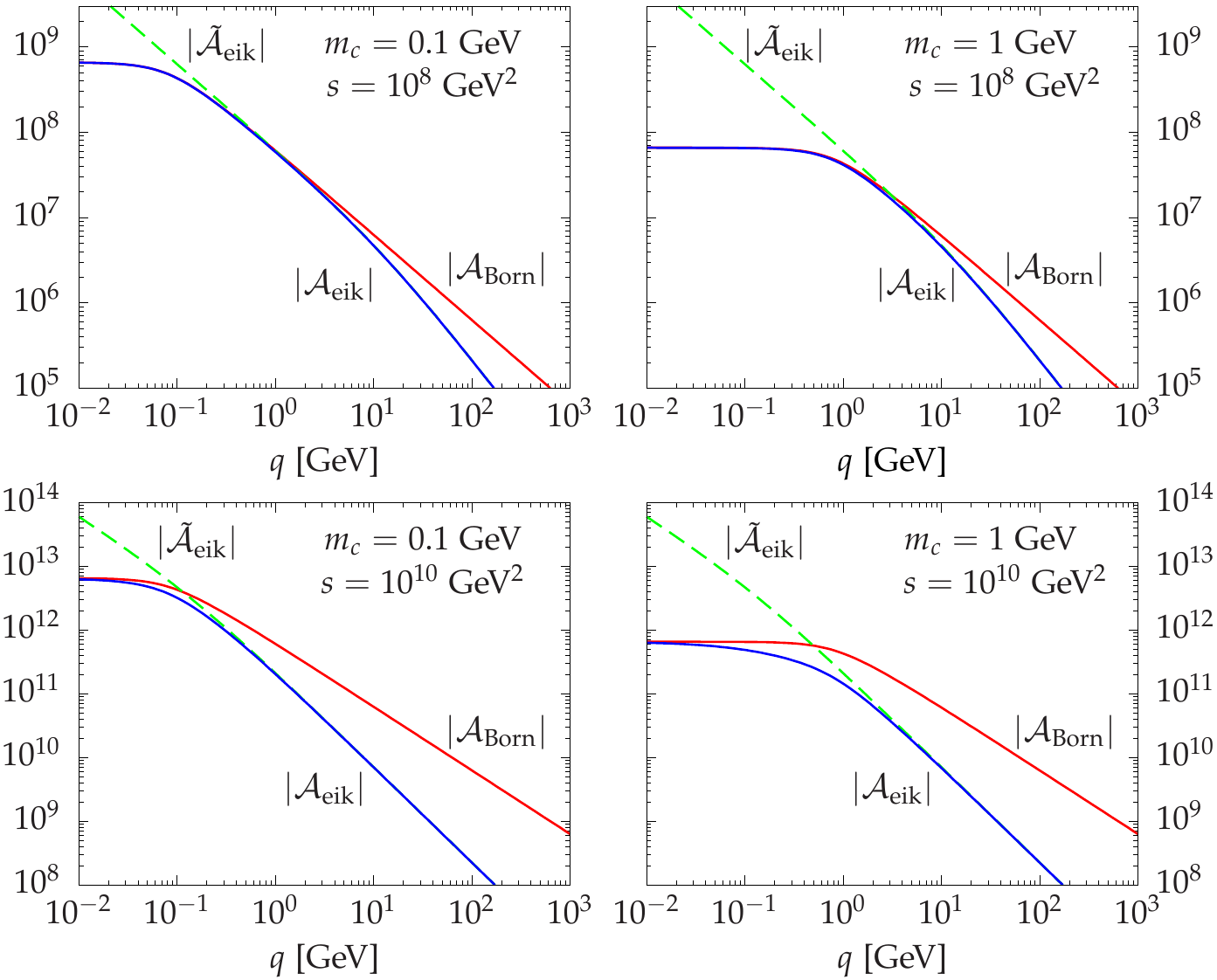}
\end{center}
\caption{Absolute value of the 
amplitudes ${\cal A}_{\rm Born}$ (in Eq.~\ref{eq5}), 
$\tilde{\cal  A}_{\rm eik}$ (in Eq.~\ref{eik2})  
and ${\cal A}_{\rm eik}$  (in Eq.~\ref{eikonal}). 
We have taken $M_5=1$ TeV with $s=(10\;{\rm TeV})^2$ (upper)
or $s=(100\;{\rm TeV})^2$ (lower) and 
$m_c=0.1$ GeV (left) or $m_c=1$ GeV (right).
}
\label{fig4}
\end{figure}
In Fig.~\ref{fig4} we provide a plot of 
the absolute value of these amplitudes. Notice that at
$q< b_c^{-1}$ the eikonal and the Born amplitudes coincide, and that 
at $q> m_c$ the correction factor goes to 1 and 
${\cal A}_{\rm eik}\approx \tilde {\cal A}_{\rm eik}$.

\subsection{Transplanckian collisions at shorter distances}
The eikonal description will  prevail as far
as the dominant impact parameter $b_s$  is larger
than the Schwarzschild radius $r_H$ of the system:
\beq
r_H(s)=M_5^{-1} \sqrt{2 \over 3\pi }
\left({s\over M_5^{2}}\right)^{1/4}.
\label{RS}
\eeq
As $b$ gets smaller, however, $q$ and the inelasticity
$y$ grow, non-linear corrections become important
and two other processes dominate: the
emission of soft gravitons (bremsstrahlung) \cite{Amati:1990xe} and the capture
of the incident particle to form a microscopic black hole 
(BH) \cite{Dimopoulos:2001hw,Giddings:2001bu,Feng:2001ib,Anchordoqui:2001ei}.
Let us discuss these processes in some more detail.

Graviton emission  appears as an imaginary contribution to the 
eikonal phase corrected by H diagrams ($\chi_H$).
This contribution is of absorptive type, it damps the elastic
cross section showing a Bloch-Nordsieck mechanism  at
work. For a given value of $b$, the
average number $N$ of gravitons radiated during the 
scattering can be read directly from $\chi_H$:
\beq
N={\rm Im}\;(\chi_H)\approx 
\left( {b_r\over b}\right)^{5},
\eeq
with
$b_r=  r_H \left( b_c /r_H \right)^{1/ 5}$.
Therefore, the typical transverse momentum radiated in the process is
$Q\approx N\, b^{-1}$. To obtain the energy lost
by the incoming particle this momentum must be boosted 
from the c.o.m. to the target rest frame.
In an eikonal scattering the dominant impact parameter 
is $b\approx b_s$. Both the number of gravitons 
$N\approx y^{5/4}(s/M_5^2)^{3/4}$ and the energy that
each one carries 
decrease at small $y$,
implying that for $y\ll 1$ the amount of 
gravitational radiation during the scattering is small.

In a collision at $b\approx r_H$ 
the incident particle will transfer a large fraction of its 
momentum to the target, changing its trajectory and 
losing to radiation a significant fraction of energy.
At these and smaller values of $b$ one expects the formation
of a microscopic BH. 
Notice that $r_H$ grows with the c.o.m. energy as $s^{1/4}$,
{\it i.e.}, the larger the energy  the larger the transverse distance
with the target that is sufficient to place the whole system inside the 
gravitational horizon. Classical (long-distance but non-perturbative) 
gravity is then all we need to describe these collisions. 
It is not that  
massive physics, like a $Z$ boson or a string excitation, are not
to be produced in these transplanckian processes 
\cite{Amati:1987wq,Cornet:2001gy}. The main
point is that all this short-distance physics occurs 
{\it inside} the BH horizon, and thus it is unable to change 
the sequence of events that we see outside.
At energies not too far from $M_5$ ({\it e.g.}, $\sqrt{s}\approx 10M_5$),
however, 
it has been shown that a number of factors
(angular momentum, charge,
geometry of the trapped surface or total radiation before the collapse)
make a precise estimate of BH production difficult.

\subsection{Neutrino-nucleon cross section}
The model under study has then two unrelated
parameters: the scale $M_5\approx$ TeV where gravity becomes strong
and the mass $m_c$ of the first KK excitation \cite{Illana:2014bda}. 
We have learned in the previous
subsection that the second parameter is only relevant in long-distance
interactions, in particular,
it can be used to suppress the soft contributions of $q < m_c$. 

\begin{figure}
\begin{center}
\includegraphics[scale=1.]{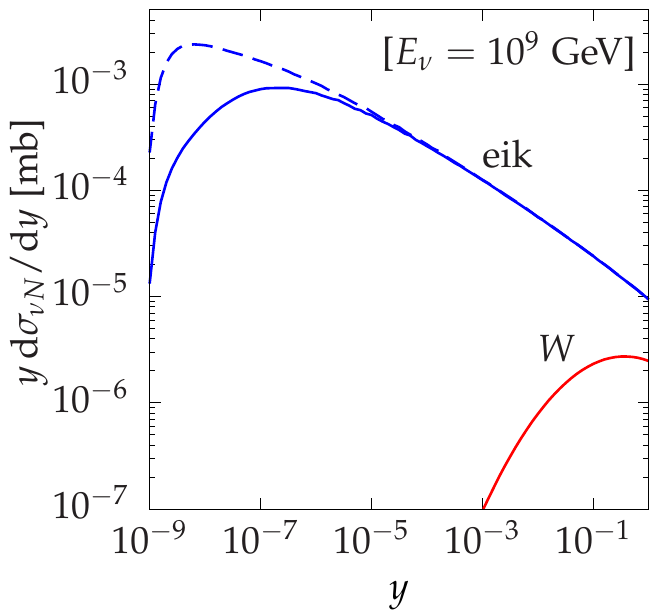}
\includegraphics[scale=1.]{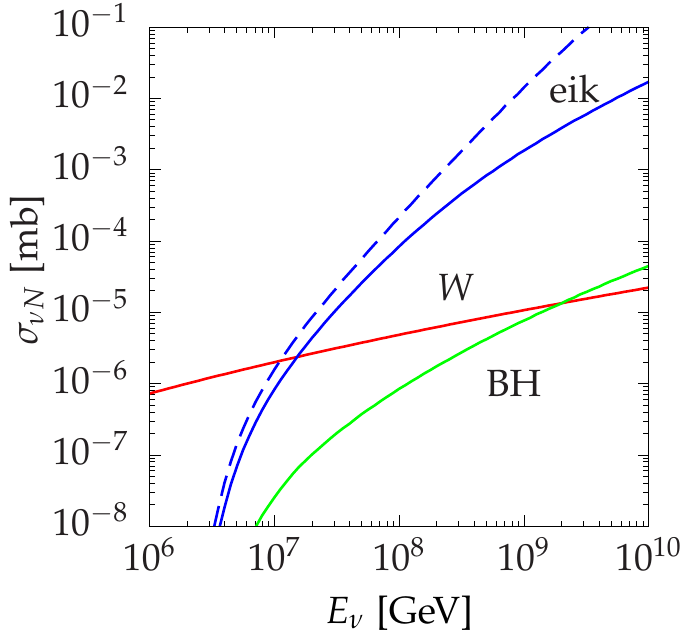}
\end{center}
\caption{Differential $\nu N$ cross section for $M_5=2$ TeV and 
$m_c=0.5,5$ GeV (dashes and solid, respectively). In the right plot
we compare the total eikonal cross section with the cross section 
for BH production and the SM cross section through $W$ exchange.
}
\label{fig5}
\end{figure}

Let us now consider a collision of a neutrino of energy $E=10^9$ GeV 
with a nucleus at rest for $M_5=2$ TeV.
At very low momentum transfer, $q\le r_p^{-1}$ with 
$r_p\approx 1/(0.2\;{\rm GeV})$ the proton radius,
the neutrino may interact coherently with a nucleon, and at even smaller 
values of $q$ it may do it with the whole nucleus. 
It is easy to see, however, that such collisions imply a very low inelasticity 
$y \le 2.7\times 10^{-11}$ ({\it i.e.,} energy depositions below
20 MeV) and a cross section 
\beq
{{\rm d} \sigma_{\rm eik}\over {\rm d} q^2} = {1\over 16\pi s^2}\,
| {\cal A}_{\rm eik} |^2
\eeq
of order $0.1$ $\mu$b. 
Therefore, the main effect will appear at 
shorter distances, when the neutrinos exchange momenta $q>1$ GeV
with the partons inside a nucleon.  The 
differential $\nu N$ cross section is then
\beq
\frac{{\rm d}\sigma_{\rm eik}}{{\rm d}y}=\int^1_{M^2_5/s} \!\! {\rm d}x\,
{1\over 16\pi x s}\, | {\cal A}_{\rm eik}(xs,y)|^2
\sum_{i=q,\bar{q},g}f_i(x,\mu)\,,
\eeq
where 
$y=q^2/(xs)$, we restrict to transplanckian collisions 
($xs\ge M_5^2$), and the PDFs $f_i(x,\mu)$ must be calculated 
at $\mu= b_s^{-1}$ for
$q>b_c^{-1}$ and $\mu=q$ when $q<b_c^{-1}$. 
Notice that
quarks and gluons interact with the same amplitude. In Fig.~\ref{fig5}
we plot this cross section for $E=10^9$ GeV, $M_5=2$ TeV and
$m_c=0.5,5$ GeV.

As for BH production in neutrino--parton interactions, 
the cross section  (also in Fig.~\ref{fig5})
can be estimated as
\beq
\sigma_{\rm BH}=\int^1_{M^2_5/s} \!\! {\rm d}x\,
\pi\, r_H^2
\sum_{i=q,\bar{q},g}f_i(x,\mu)\,,
\eeq
with $r_H(xs)$ given in Eq.~(\ref{RS}) and $\mu=r_H^{-1}$.
The eikonal process is
dominated by much larger impact parameters, 
and its 
overlapping with the (inclusive)
geometrical cross section $\sigma_{\rm BH}$ will be negligible. 
The BH, of mass $M_{BH}\approx \sqrt {xs}$ and
temperature $T=1/(2\pi r_H)$, will evaporate 
almost instantly into SM particles
(see \cite{Draggiotis:2008jz} and references therein).

In summary, TeV gravity 
implies a $\nu N$ cross section that
grows fast with the energy above the threshold 
$E_{\rm th}\approx M_5^2/(2 m_p)$ defining the transplanckian regime.
In particular, the collision of an ultrahigh energy neutrino 
with the partons inside the nucleon may give events 
where the neutrino deposits a small fraction $y=10^{-7}$--$10^{-3}$ of
its energy and keeps going. The cross section $\sigma_{\rm eik}$
for these collisions
can be well above the standard one mediated by the  $W$ and $Z$ 
gauge bosons,
and will always be larger than the shorter distance  
interactions producing a microscopic BH.

\section{Signal at neutrino telescopes}
What are the experimental bounds on this TeV gravity set up? Let us 
briefly discuss cosmological \cite{Hannestad:2001nq}, 
astrophysical \cite{Hannestad:2003yd}, 
collider \cite{Franceschini:2011wr} and cosmic ray
observations; we will argue that the scenario under study 
may be out of reach
everywhere except for at neutrino telescopes.

In this model  the KK gravitons have 
a relatively short lifetime before they decay into gammas ($n_V=1$) and 
light fermions ($u,d,e,\mu,\nu$:  $n_f=19$); for the first excitation 
($m_1={5\over 4}\pi k$)
we find \cite{Han:1998sg}:
\beq
\tau\approx {50 \pi\over 4 n_V+n_f} \;{M_5^3\over m_1^4} = 4\times 
10^{-14} \;{\rm s} \; \left({1\; {\rm GeV}\over m_1}\right)^4
\left({M_5\over 2\; {\rm TeV}}\right)^3\,.
\eeq
Due to their increased coupling to matter (relative to the massless 
graviton), 
in the early universe they may be in thermal equilibrium 
at temperatures $T\ge m_c$. If their mass is larger than 10 MeV, 
however, at the time of primordial
nucleosynthesis all of them will be gone and their
decay products thermalized: these 
massive gravitons are then cosmologically safe.

In astrophysics, if their mass is below 100 MeV they may be produced 
abundantly in protoneutron stars during a core collapse. 
For $m_c=50$--$100$ MeV and $M_5\ge 2$ TeV the
short lifetime of these gravitons does not let them escape the core 
and change significantly
the dynamics of the explosion. In particular, they would not
shorten the duration of the neutrino signal produced during 
a supernova explosion.
At lifetimes $\tau \approx 10^{-4}$ s these massive gravitons 
could even play a role in the
revival of the stalled shock front in a core 
collapse \cite{Fuller:2009zz,Albertus:2015xra,Rembiasz:2018lok}.

In collider experiments they may introduce displaced vertices
or rare decays ({\it e.g.}, $K\to \pi g^{(n)}$), 
but their production cross section is always
suppressed by inverse powers of $M_5$ and several TeV
may suffice to evade all constraints.

Finally, there may also be bounds from
air shower experiments. Ultrahigh energy cosmic rays have 
strong interactions with matter, so gravity may only introduce 
order 1 corrections that do not seem excluded. We are left 
with airs showers produced by high energy neutrinos.
The AUGER observatory, in particular, looks for inclined 
events that start deep into the atmosphere, setting constraints on
a neutrino flux at $10^8$--$10^{11}$ GeV \cite{Abreu:2013zbq}. 
Strong 
TeV gravity will not change the neutrino interactions at $E_\nu < 1$ PeV, but
it could multiply by hundred the cross section of cosmogenic neutrinos
at higher energies.
An obvious observation is then, would this 
large cross section introduce a signal in air shower
experiments detectable at AUGER?
Not
necessarily. The reason is that, although large, the 
gravitational cross section 
is very soft: a $10^9$ GeV
neutrino will typically start a TeV--PeV atmospheric shower, which is
below the energy threshold at AUGER. The same argument would apply to 
ANITA or LUNASKA: they search for a single $10^{10}$ GeV 
energy deposition from a cosmogenic neutrino, but the 
typical inelasticity in an eikonal collision is just 
$y\approx 10^{-5}$.

A cosmogenic neutrino  
would leave a very characteristic signal in a 
km$^3$ telescope \cite{Illana:2014bda}; 
let us briefly discuss its main features. 

\begin{itemize}

\item First, as we have already mentioned, the typical energy of 
an event is {\it not} the energy of the cosmogenic
neutrino; eikonal collisions are very soft and  translate
into TeV--PeV energy depositions. 

\item Second, the signal will always come from 
downgoing or near-horizontal directions, never from
upgoing directions. The reason is that
$10^8$--$10^{11}$ GeV neutrinos are unable to cross the 
Earth. In Fig.~\ref{fig6} we plot the probability that
a high-energy neutrino reaches IceCube from different 
zenith angles without experiencing a standard interaction 
with matter. Hard gravitational interactions, which
may reduce further the reach of these neutrinos,
have not been included.

\begin{figure}
\begin{center}
{\includegraphics[scale=1.]{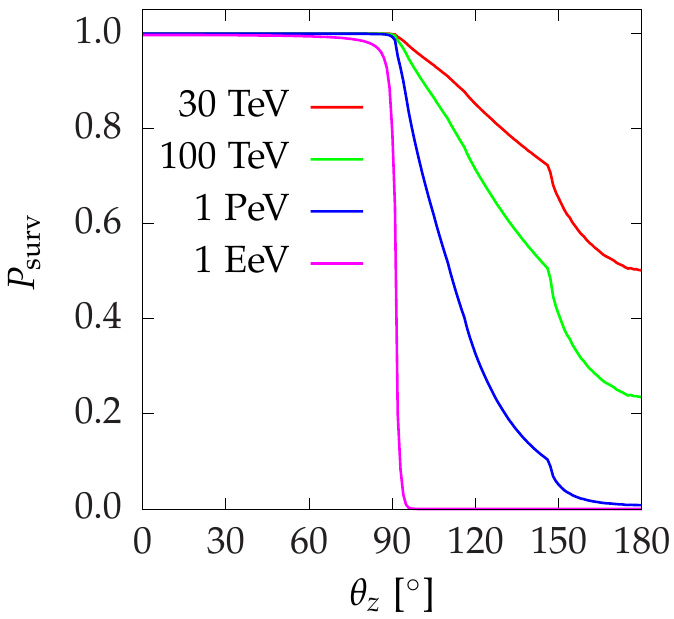}}
\end{center}
\caption{Probability $P_{\rm surv}$ that a neutrino reaches IceCube from
a zenith angle $\theta_z$ for several energies $E_\nu$ (we have used
the $\nu N$ cross section in \cite{Connolly:2011vc}).
}
\label{fig6}
\end{figure}

\item In addition, the signal will only introduce shower events, never 
tracks. Notice that the eikonal amplitude describing multigraviton 
exchange does not change the incident neutrino into a charged lepton.

\item Finally, the eikonal scattering will not {\it stop} the 
cosmogenic neutrino: after the first collision the neutrino 
keeps going with basically
the same energy and can  
interact multiple times with matter, possibly inside the
detector, 
before it has a harder (standard or gravitational)
interaction that reduces  its energy. 

\end{itemize}
The last point is specially interesting, as there are no standard
events giving {\it double bangs} of TeV--PeV energy. Tau neutrinos
may indeed produce this topology, but at much higher energies: 
we need $E_\nu\approx 10^8$ GeV 
to expect 100 meters between the creation and the decay points of a 
tau lepton. If, for example, the eikonal cross section 
is $\sigma_{\rm eik}=4$ $\mu$b, then 
the mean free path between interactions in ice would be around 
$\lambda_{\rm eik}=4$ km. It is ease to deduce that the probability
to have $N$ interactions ({\it bangs}) along a length $L$ is 
\cite{Illana:2005pu}
\beq
P_{N}(L)={\rm e}^{-{L/\lambda_{\rm eik}}}\, \frac{(L/\lambda_{\rm eik})^N}{N!}\ .
\eeq
In that case the neutrino would have a 19.5\% probability for a single
interaction when crossing 1 km of ice or a 2.4\% probability
to produce a double bang within the same distance.

\section{Summary and discussion}

In models with extra dimensions the fundamental scale $M_D$ of gravity
may take any value between $10^3$ and $10^{19}$ GeV, depending on
the details (topology, length, curvature) of the compactification
space. Due to the spin 2 of the graviton, in the transplanckian regime 
($s>M_D^2$) any collision is dominated
by classical (long-distance but non-perturbative) 
gravity, as all the short-distance physics is trapped inside 
a BH horizon. At impact parameters larger than $r_H$
one expects graviton mediated interactions with 
a large cross section but of very small inelasticity.
In particular, a cosmogenic neutrino of $10^8$--$10^{11}$ GeV could
scatter off matter and produce a TeV--PeV shower.
This scenario is actually a particular realization of a more
general one where the UV completion of the SM occurs (at the TeV 
or at a higher scale) 
through
{\it classicalization} \cite{Dvali:2010jz}.

The 
signal suggested by TeV gravity at large neutrino telescopes 
consists then of an excess of TeV--PeV shower events 
from downgoing directions. Indeed, the
IceCube's excess of 
high energy starting events \cite{Aartsen:2014gkd}
exhibits a preference for downgoing
versus upgoing directions and for showers versus tracks, so there 
may be room for this type of physics in the current data.

There are two kind of observations that could clearly favor 
these scenarios. First, double-bang events of TeV--PeV energy. 
Second, the {\it absence} of cosmogenic neutrinos. If 
$10^8$--$10^{11}$ GeV neutrinos do not appear in current and
future searches, the reason could
be that at those ultrahigh energies they do not look like neutrinos
anymore. Instead of an invisible particle able to penetrate 2 km
of ice and deposit $10^{10}$ GeV, TeV gravity could turn them into
a particle that interacts
frequently ({\it e.g.}, every 0.1--1 km of ice) but deposits a much smaller
amount of energy. Its detection may then require 
a different strategy.
At any rate, these are two questions that the next generation
of neutrino telescopes should be able to answer.

\section*{Acknowledgments}
This work has been supported by MICINN of Spain 
(FPA2016-78220, RED2018-102661-T, FPA2017-90566-REDC) 
and by Junta de Andaluc\'\i a (SOMM17/6104/UGR and FQM101).


\begin{thebibliography}{99} 

\bibitem{ArkaniHamed:1998rs}
  N.~Arkani-Hamed, S.~Dimopoulos and G.~R.~Dvali,
  Phys.\ Lett.\ B {\bf 429} (1998) 263
  [hep-ph/9803315].

\bibitem{Antoniadis:1990ew}
  I.~Antoniadis,
  Phys.\ Lett.\ B {\bf 246} (1990) 377.

\bibitem{Horava:1996ma}
  P.~Horava and E.~Witten,
  Nucl.\ Phys.\ B {\bf 475} (1996) 94
  [hep-th/9603142].

\bibitem{Polchinski:1998rr}
  J.~Polchinski,
{\it String theory. Vol. 2: Superstring theory and beyond},
Cambridge University Press (2007).

\bibitem{Randall:1999ee}
  L.~Randall and R.~Sundrum,
  Phys.\ Rev.\ Lett.\  {\bf 83} (1999) 3370
  [hep-ph/9905221].

\bibitem{Gherghetta:2010cj}
  T.~Gherghetta,
  arXiv:1008.2570 [hep-ph].

\bibitem{Maldacena:1997re}
  J.~M.~Maldacena,
  Int.\ J.\ Theor.\ Phys.\  {\bf 38} (1999) 1113
   [Adv.\ Theor.\ Math.\ Phys.\  {\bf 2} (1998) 231]
  [hep-th/9711200].

\bibitem{ArkaniHamed:2000ds}
  N.~Arkani-Hamed, M.~Porrati and L.~Randall,
  JHEP {\bf 0108} (2001) 017
  [hep-th/0012148].

\bibitem{Dienes:1998sb}
  K.~R.~Dienes, E.~Dudas and T.~Gherghetta,
  Nucl.\ Phys.\ B {\bf 557} (1999) 25
  [hep-ph/9811428].

\bibitem{Barcelo:2011vk}
  R.~Barcelo, A.~Carmona, M.~Masip and J.~Santiago,
  Phys.\ Lett.\ B {\bf 707} (2012) 88
  [arXiv:1106.4054 [hep-ph]].

\bibitem{Semikoz:2003wv}
  D.~V.~Semikoz and G.~Sigl,
  JCAP {\bf 0404} (2004) 003
  [hep-ph/0309328].

\bibitem{Ahlers:2012rz}
  M.~Ahlers and F.~Halzen,
  Phys.\ Rev.\ D {\bf 86} (2012) 083010
  [arXiv:1208.4181 [astro-ph.HE]].

\bibitem{Fodor:2003ph}
  Z.~Fodor, S.~D.~Katz, A.~Ringwald and H.~Tu,
  JCAP {\bf 0311} (2003) 015
  [hep-ph/0309171].

\bibitem{Ahlers:2010fw}
  M.~Ahlers, L.~A.~Anchordoqui, M.~C.~Gonzalez-Garcia, F.~Halzen and S.~Sarkar,
  Astropart.\ Phys.\  {\bf 34} (2010) 106
  [arXiv:1005.2620 [astro-ph.HE]].

\bibitem{Deaconu:2017eyy}
  C.~Deaconu [ANITA Collaboration],
  EPJ Web Conf.\  {\bf 135} (2017) 01008.

\bibitem{James:2009rc}
  C.~W.~James, R.~J.~Protheroe, R.~D.~Ekers, J.~Alvarez-Muniz, R.~A.~McFadden, C.~J.~Phillips, P.~Roberts and J.~D.~Bray,
  Mon.\ Not.\ Roy.\ Astron.\ Soc.\  {\bf 410} (2011) 885
  [arXiv:0906.3766 [astro-ph.HE]].

\bibitem{Aartsen:2014njl}
  M.~G.~Aartsen {\it et al.} [IceCube Collaboration],
{\it IceCube-Gen2: A Vision for the Future of Neutrino Astronomy in Antarctica},
  arXiv:1412.5106 [astro-ph.HE].

\bibitem{Aartsen:2017mau}
  M.~G.~Aartsen {\it et al.} [IceCube Collaboration],
{\it The IceCube Neutrino Observatory - Contributions to ICRC 2017 Part II: Properties of the Atmospheric and Astrophysical Neutrino Flux},
  arXiv:1710.01191 [astro-ph.HE].

\bibitem{Giudice:1998ck}
  G.~F.~Giudice, R.~Rattazzi and J.~D.~Wells,
  Nucl.\ Phys.\ B {\bf 544} (1999) 3
  [hep-ph/9811291].

\bibitem{Giudice:2004mg}
  G.~F.~Giudice, T.~Plehn and A.~Strumia,
  Nucl.\ Phys.\ B {\bf 706} (2005) 455
  [hep-ph/0408320].

\bibitem{Amati:1993tb}
  D.~Amati, M.~Ciafaloni and G.~Veneziano,
  Nucl.\ Phys.\ B {\bf 403} (1993) 707.

\bibitem{Emparan:2001ce}
  R.~Emparan,
  Phys.\ Rev.\ D {\bf 64} (2001) 024025
  [hep-th/0104009].

\bibitem{Giudice:2001ce}
  G.~F.~Giudice, R.~Rattazzi and J.~D.~Wells,
  Nucl.\ Phys.\ B {\bf 630} (2002) 293
  [hep-ph/0112161].

\bibitem{Emparan:2001kf}
  R.~Emparan, M.~Masip and R.~Rattazzi,
  Phys.\ Rev.\ D {\bf 65} (2002) 064023
  [hep-ph/0109287].

\bibitem{Illana:2004qc}
  J.~I.~Illana, M.~Masip and D.~Meloni,
  Phys.\ Rev.\ Lett.\  {\bf 93} (2004) 151102
  [hep-ph/0402279].

\bibitem{Illana:2005pu}
  J.~I.~Illana, M.~Masip and D.~Meloni,
  Phys.\ Rev.\ D {\bf 72} (2005) 024003
  [hep-ph/0504234].

\bibitem{Amati:1990xe}
  D.~Amati, M.~Ciafaloni and G.~Veneziano,
  Nucl.\ Phys.\ B {\bf 347} (1990) 550.

\bibitem{Dimopoulos:2001hw}
  S.~Dimopoulos and G.~L.~Landsberg,
  Phys.\ Rev.\ Lett.\  {\bf 87} (2001) 161602
  [hep-ph/0106295].

\bibitem{Giddings:2001bu}
  S.~B.~Giddings and S.~D.~Thomas,
  Phys.\ Rev.\ D {\bf 65} (2002) 056010
  [hep-ph/0106219].

\bibitem{Feng:2001ib}
  J.~L.~Feng and A.~D.~Shapere,
  Phys.\ Rev.\ Lett.\  {\bf 88} (2002) 021303
  [hep-ph/0109106].

\bibitem{Anchordoqui:2001ei}
  L.~Anchordoqui and H.~Goldberg,
  Phys.\ Rev.\ D {\bf 65} (2002) 047502
  [hep-ph/0109242].

\bibitem{Amati:1987wq}
  D.~Amati, M.~Ciafaloni and G.~Veneziano,
  Phys.\ Lett.\ B {\bf 197} (1987) 81.

\bibitem{Cornet:2001gy}
  F.~Cornet, J.~I.~Illana and M.~Masip,
  Phys.\ Rev.\ Lett.\  {\bf 86} (2001) 4235
  [hep-ph/0102065].

\bibitem{Illana:2014bda}
  J.~I.~Illana, M.~Masip and D.~Meloni,
  Astropart.\ Phys.\  {\bf 65} (2015) 64
  [arXiv:1410.3208 [hep-ph]].

\bibitem{Draggiotis:2008jz}
  P.~Draggiotis, M.~Masip and I.~Mastromatteo,
  JCAP {\bf 0807} (2008) 014
  [arXiv:0805.1344 [hep-ph]].

\bibitem{Hannestad:2001nq}
  S.~Hannestad,
  Phys.\ Rev.\ D {\bf 64} (2001) 023515
  [hep-ph/0102290].

\bibitem{Hannestad:2003yd}
  S.~Hannestad and G.~G.~Raffelt,
  Phys.\ Rev.\ D {\bf 67} (2003) 125008
   Erratum: [Phys.\ Rev.\ D {\bf 69} (2004) 029901]
  [hep-ph/0304029].

\bibitem{Franceschini:2011wr}
  R.~Franceschini, P.~P.~Giardino, G.~F.~Giudice, P.~Lodone and A.~Strumia,
  JHEP {\bf 1105} (2011) 092
  [arXiv:1101.4919 [hep-ph]].

\bibitem{Han:1998sg}
  T.~Han, J.~D.~Lykken and R.~J.~Zhang,
  Phys.\ Rev.\ D {\bf 59} (1999) 105006
  [hep-ph/9811350].

\bibitem{Fuller:2009zz}
  G.~M.~Fuller, A.~Kusenko and K.~Petraki,
  Phys.\ Lett.\ B {\bf 670} (2009) 281
  [arXiv:0806.4273 [astro-ph]].

\bibitem{Albertus:2015xra}
  C.~Albertus, M.~Masip and M.~A.~Pérez-García,
  Phys.\ Lett.\ B {\bf 751} (2015) 209
  [arXiv:1509.03306 [astro-ph.HE]].

\bibitem{Rembiasz:2018lok}
  T.~Rembiasz, M.~Obergaulinger, M.~Masip, M.~Á.~Pérez-García, M.~Á.~Aloy and C.~Albertus,
  Phys.\ Rev.\ D {\bf 98} (2018) no.10,  103010
  [arXiv:1806.03300 [astro-ph.HE]].

\bibitem{Abreu:2013zbq}
  P.~Abreu {\it et al.} [Pierre Auger Collaboration],
  Adv.\ High Energy Phys.\  {\bf 2013} (2013) 708680
  [arXiv:1304.1630 [astro-ph.HE]].

\bibitem{Connolly:2011vc}
  A.~Connolly, R.~S.~Thorne and D.~Waters,
  Phys.\ Rev.\ D {\bf 83} (2011) 113009
  [arXiv:1102.0691 [hep-ph]].

\bibitem{Dvali:2010jz}
  G.~Dvali, G.~F.~Giudice, C.~Gomez and A.~Kehagias,
  JHEP {\bf 1108} (2011) 108
  [arXiv:1010.1415 [hep-ph]].

\bibitem{Aartsen:2014gkd}
  M.~G.~Aartsen {\it et al.} [IceCube Collaboration],
  Phys.\ Rev.\ Lett.\  {\bf 113} (2014) 101101
  [arXiv:1405.5303 [astro-ph.HE]].


\vfil
\end{thebibliography}
\end{document}